\documentclass[aps,prl,superscriptaddress,twocolumn,nobibnotes]{revtex4-2}

\usepackage{graphicx}
\usepackage{dcolumn}
\usepackage{bm}
\usepackage{natbib}
\usepackage{amsmath}
\usepackage{bibunits}

\begin{document}

\title{Spin Polarization of a Two-Dimensional Electron Liquid}

\author{G.~A.~Nikolaev}
\email{nikolaevgk@gmail.com}
\affiliation{Osipyan Institute of Solid State Physics, Russian Academy of Sciences, Chernogolovka, Moscow Region 142432, Russia}
\affiliation{Moscow Institute of Physics and Technology, Dolgoprudny, Moscow Region 141701, Russia}

\author{M.~E.~Sergeev}
\affiliation{Osipyan Institute of Solid State Physics, Russian Academy of Sciences, Chernogolovka, Moscow Region 142432, Russia}
\affiliation{Moscow Institute of Physics and Technology, Dolgoprudny, Moscow Region 141701, Russia}

\author{I.~V.~Kukushkin}
\affiliation{Osipyan Institute of Solid State Physics, Russian Academy of Sciences, Chernogolovka, Moscow Region 142432, Russia}

\author{A.~V.~Shchepetilnikov}
\affiliation{Osipyan Institute of Solid State Physics, Russian Academy of Sciences, Chernogolovka, Moscow Region 142432, Russia}
\affiliation{National Research University Higher School of Economics, 101000 Moscow, Russia}

\begin{abstract}
We show experimentally that the longitudinal resistance of a strongly correlated two-dimensional electron system provides a
quantitative measure of its spin polarization. Using electrically detected electron spin resonance to independently
calibrate the spin state, we establish a parameter-free relation between magnetotransport and spin polarization,
allowing the latter to be determined from transport alone. This approach enables the magnetic state of the electron
liquid to be mapped across a broad range of carrier densities and magnetic fields.
\end{abstract}

\maketitle
\begin{bibunit}[apsrev4-2]

Two-dimensional electron liquids provide a unique platform for exploring how electron--electron interactions couple
charge transport to the spin degrees of freedom. Understanding this interplay is a central challenge in modern condensed-matter physics,
as spin-dependent transport not only provides insight into the many-body state of correlated electrons but also offers new opportunities
for spin-functional electronic devices~\cite{Tokura-2017-Nature}. Recent studies have shown that Coulomb interactions can give rise to spin-selective
conductivity~\cite{Shih-2025-Nature} and spontaneous spin--valley polarization~\cite{Boddison-2025-Science} in atomically thin semiconductors,
while magnetotransport measurements have revealed strongly enhanced spin susceptibility and emergent magnetic
correlations in the vicinity of interaction-driven electronic phases~\cite{Falson-2022-NatureMaterials,YangXu-2025-Arxiv}.
These advances establish spin-dependent transport as a powerful probe of correlated quantum matter.
In the present work, we obtain and experimentally validate a direct quantitative relation between electrical transport
and the spin polarization of a two-dimensional electron liquid.

The key quantity characterizing the magnetic state of a two-dimensional electron system is the spin polarization,
which is defined by the population imbalance between the Zeeman-split spin subbands,
\begin{equation}
\label{eq:xi}
\xi=\frac{n_{\uparrow}-n_{\downarrow}}{n},
\end{equation}
where $n_{\uparrow}$ and $n_{\downarrow}$ are the populations of the lower- and upper-energy spin subbands, respectively,
and $n=n_{\uparrow}+n_{\downarrow}$ is the total electron density.
As the Zeeman energy increases, electrons are progressively transferred to the lower spin subband until,
at the critical field $B_c$, the Zeeman energy equals the nondegenerate Fermi energy and the system becomes
fully spin polarized~\cite{Vitkalov-2000-PRL,Okamoto-1999-PRL},
\begin{equation}
\label{eq:Bc}
B_c(n)=\left(\frac{2\pi\hbar^2}{\mu_B}\right)\frac{n}{g^*m^*},
\end{equation}
where $n$ is the electron density, $\mu_B$ is the Bohr magneton, $g^*$ and $m^*$ are the effective Landé $g$-factor
and electron mass, respectively. As the parallel magnetic field spin-polarizes the electron system, screening of the disorder potential becomes less effective,
leading to enhanced impurity scattering and an increase in the resistance~\cite{Dolgopolov-Gold-2000-JETPLetters}.
Accordingly, $B_c$ is identified from a pronounced change in the parallel-field magnetoresistance and is widely used as a signature of
complete spin polarization in both conventional semiconductor and van der Waals heterostructures~\cite{YangXu-2025-Arxiv, Falson-2022-NatureMaterials, Spivak-2010-RMP}.
Within this framework, electron--electron interactions are incorporated through the interaction-induced enhancement of $g^* m^*(n)$,
thereby reducing $B_c$. However, many-body interactions also drive a nonlinear evolution of the spin polarization with magnetic field~\cite{Tutuc-2003-PRB},
in contrast to the linear dependence expected for a non-interacting Fermi gas.
Despite the widespread use of parallel-field magnetotransport as an indirect probe of spin polarization, its role has
remained largely limited to identifying the onset of complete spin polarization through $B_c$. A direct quantitative relation
between the magnetoresistance itself and the spin state of the interacting electron liquid has so far remained unknown.

Electron spin resonance (ESR), the most direct spectroscopic probe of electronic spin states, provides a natural route
to address this problem. Originally developed for the study of paramagnetic species, ESR has evolved into a versatile platform
spanning condensed-matter physics, chemistry, biology, quantum sensing, molecular spin systems, and spin-based quantum
information science~\cite{Shchepetilnikov-2025-UFN,Lazarova-2021-AnalChem,Zhang-2021-ACS,Weiss-2017-Nature,Polash-2023-AIP, Grigoryan-2026-PRB}.
Driven by rapid advances in microwave and sub-terahertz technologies, ESR is experiencing renewed momentum, with major
progress in spectroscopy of low-dimensional materials, correlated electron systems, and
quantum devices~\cite{Feder-2025-Nature,Sellies-2023-Nature,Mena-2024-PRL, Zeisner-2020-PRMaterials, Senyk-2023-PRMaterials,Morissette-2023-Nature,Shchepetilnikov-2024-PRL}.

In this work, we establish a direct, parameter-free relation between the longitudinal resistance and the equilibrium
spin polarization of a strongly correlated two-dimensional electron system (two-dimensional electron liquid).
Electrically detected ESR serves as an independent probe of the spin state to validate this relation.
Once validated, the equilibrium spin polarization can be determined from conventional magnetotransport
measurements alone over a broad range of electron densities and in-plane magnetic fields, providing a
simple experimental route to mapping the evolution from a paramagnetic to a fully spin-polarized Fermi liquid.

\begin{figure*}
\includegraphics[width=2\columnwidth,clip]{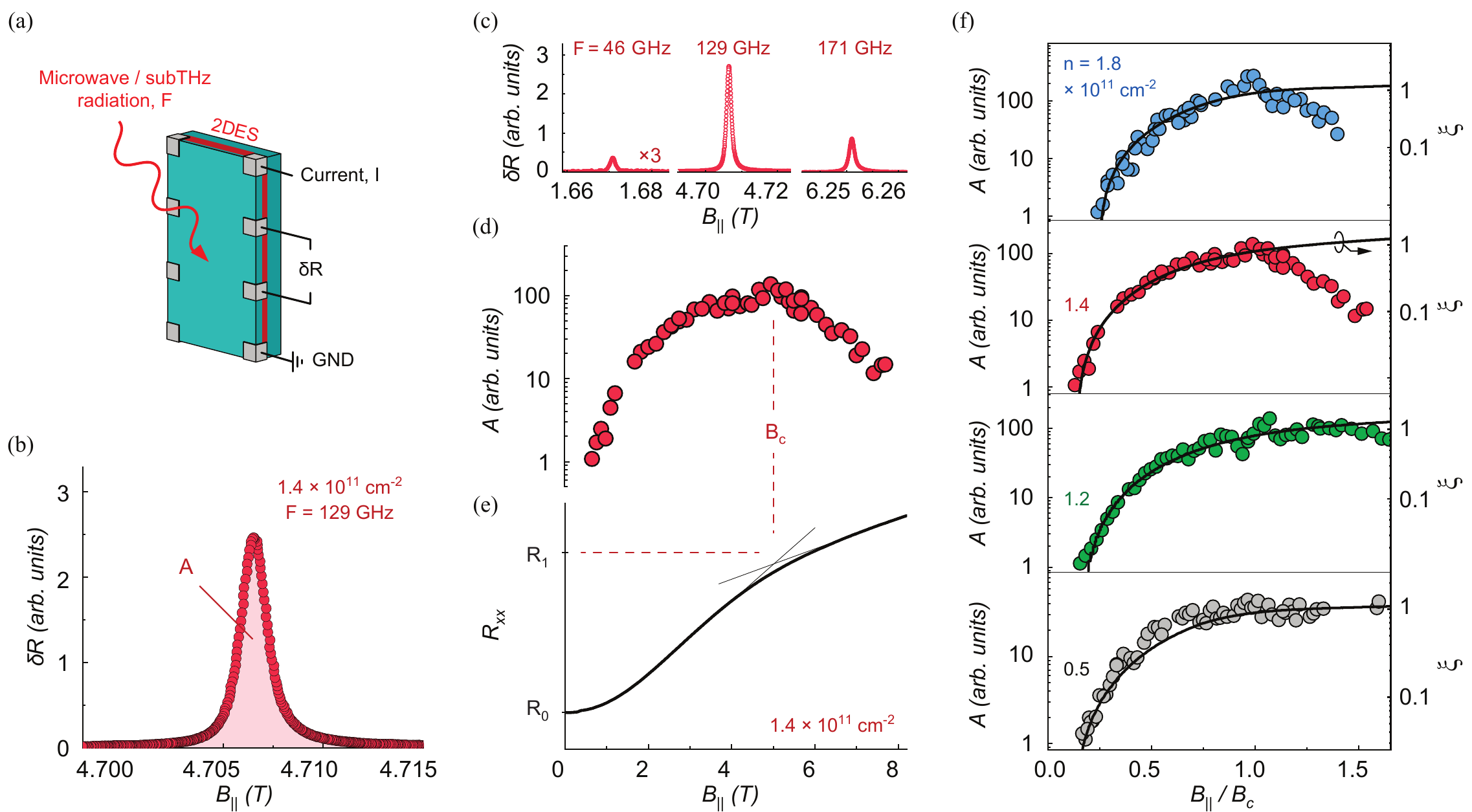}
\caption{
    (a) Schematic of the experimental configuration for broadband electrically detected ESR. Microwave or sub-terahertz
    radiation induces a resonant change in the longitudinal resistance, $\delta R$, which is detected electrically.
    (b) Representative ESR resonance measured at $F=129$~GHz for the sample with electron density $n=1.4\times10^{11}$~cm$^{-2}$.
    The shaded area defines the integrated ESR intensity used throughout this work to minimize the influence of
    inhomogeneous line broadening.
    (c) ESR spectra measured at three excitation frequencies, corresponding
    to different resonance magnetic fields. The resonance intensity changes by nearly two orders of magnitude despite
    comparable linewidths.
    (d) Integrated ESR intensity as a function of parallel magnetic field. The intensity
    increases rapidly with magnetic field, reaches a maximum close to the critical field $B_c$, and decreases at higher
    fields.
    (e) Parallel-field magnetoresistance of the same sample. The critical field $B_c$, marking complete
    spin polarization, is determined from the intersection of the low- and high-field asymptotes.
    (f) Integrated ESR intensity for all investigated samples plotted as a function of normalized magnetic field $B_{\parallel}/B_c$.
    Solid curves show the spin polarization reconstructed independently from magnetotransport using Eq.~(\ref{eq:5}).
    The agreement between the data and the reconstructed curves verifies Eq.~(\ref{eq:5}) and demonstrates that the magnetic-field dependence of the ESR intensity is governed primarily
    by the evolution of the spin polarization, while deviations at $B_{\parallel}>B_c$ reflect the increasing influence of orbital effects.
}\label{fig1}
\end{figure*}

The emergence of atomically thin materials, including graphene, transition-metal dichalcogenides, and other van der Waals heterostructures,
has created new opportunities for exploring strongly interacting electrons in reduced dimensions. However,
their microscopic spin properties remain difficult to probe by electron spin resonance because of the intrinsically weak
microwave absorption of layered materials. Consequently, ESR has so far been realized only under exceptional conditions,
for example in magic-angle twisted bilayer graphene~\cite{Morissette-2023-Nature}.
Other van der Waals heterostructures generally exhibit only a weak microwave response under typical excitation
powers~\cite{Mani-2012-NatureComm, Anlauf-2021-PRB,Dinar-2025-PRApplied, Morissette-2023-Nature}.
Conventional semiconductor heterostructures provide a versatile platform for studying correlated two-dimensional electrons.
In particular, high-mobility ZnO/MgZnO heterostructures exhibit the fractional quantum Hall effect~\cite{Tsukazaki-2010-Nature,Falson-2015-Nature, Falson-2018-IOP}
and strongly correlated phases~\cite{Falson-2018-ScienceAdv,Falson-2022-NatureMaterials},
and enable optical~\cite{Solovyev-2015-AIP,Solovyev-2017-PRB,Vankov-2022-PRB,Berezhnoy-2024-PRB}
and microwave probes, including resistance oscillations~\cite{Shchepetilnikov-2019-PRB},
magnetoplasmon resonance~\cite{Kozlov-2015-PRB},
and electrically detected spin resonance~\cite{Kozuka-2014-PRB,Shchepetilnikov-2021-PRB,Shchepetilnikov-2023-PRB,Shchepetilnikov-2024-JETPLetters}.

Spin resonance was detected through the microwave-induced variation of the longitudinal resistance,
$\delta R$, using the configuration illustrated in Fig.~\ref{fig1}a. Broadband microwave excitation
($1$--$300$~GHz) was provided by custom-built waveguide and coaxial-line probe assemblies. The transport
response was recorded using a double lock-in scheme to increase signal-to-noise ratio~\cite{Shchepetilnikov-2021-PRB, Shchepetilnikov-2024-PRL}.
A representative ESR line measured in a ZnO/MgZnO two-dimensional electron system
with $n=1.4\times10^{11}$~cm$^{-2}$ at $F=129$~GHz is shown in Fig.~\ref{fig1}b.
Because spatial inhomogeneity can substantially
modify the resonance profile~\cite{Nikolaev-2026-APL}, we characterize the response by the integrated amplitude $A(B, T)$
evaluated after applying standard ESR procedures for radiation-intensity normalization and
background subtraction~\cite{Shchepetilnikov-2023-PRB, Shchepetilnikov-2024-PRL} (for more details, see Supplemental Material~II).
Throughout the temperature and magnetic-field range considered here, $k_{\mathrm B}T$
remains much smaller than both the Zeeman and Fermi energies, rendering the equilibrium spin polarization effectively temperature independent.

Figure~\ref{fig1}c compares ESR spectra acquired at $F=46$, $129$, and $171$~GHz at $T=1.5$~K for the sample with $n=1.4\times10^{11}$~cm$^{-2}$.
The integrated ESR amplitude varies by nearly two orders of magnitude, revealing a strong dependence of the electrically
detected response on the electronic state.
The complete field dependence is presented in Fig.~\ref{fig1}d: the increase in ESR amplitude $A(B, T)$ follows the progressive alignment
of electron spins. Above $B_c$, the spin polarization is saturated, and the subsequent suppression of the signal arises from a spin-polarization-independent mechanism.
This maximum coincides with the kink in the parallel-field magnetoresistance shown in Fig.~\ref{fig1}e.
The corresponding critical field $B_c$ is determined from the intersection of linear fits to the magnetoresistance
on either side of the kink~\cite{Vitkalov-Klapwijk-2001-PRL,YangXu-2025-Arxiv}.

The same behavior is observed across the sample series. Figure~\ref{fig1}f
shows the ESR amplitude versus $B_{\parallel}/B_c$ for $n=0.5$, $1.2$, $1.4$, and $1.8\times10^{11}$~cm$^{-2}$,
with corresponding $B_c=1.1$, $4.2$, $5.0$, and $7.6$~T, respectively.
In all samples, the ESR amplitude rises rapidly as the system approaches complete spin polarization ($B_{\parallel}/B_c\simeq1$).
Beyond $B_c$, however, the behavior becomes density dependent: the amplitude decreases markedly in the two highest-density
samples, more weakly at $n=1.2\times10^{11}$~cm$^{-2}$, and shows essentially no suppression at $n=0.5\times10^{11}$~cm$^{-2}$.
We attribute this high-field suppression to magneto-orbital coupling arising from the finite thickness of the electron layer,
which modifies both magnetotransport~\cite{Tutuc-2003-PRB,DasSarma-2000-PRL} and the efficiency of electron excitation~\cite{Kulik-2000-PRB}.
The systematic enhancement of the suppression with increasing electron density supports this interpretation,
as larger $B_c$ shifts the measurements to higher magnetic fields where orbital effects become progressively stronger.
A detailed analysis of this decrease is beyond the scope of the present manuscript.

To identify the ESR detection mechanism and establish the relation between spin polarization and longitudinal resistance,
we focus on the lowest-density sample,
$n=0.5\times10^{11}$~cm$^{-2}$, for which the small value of $B_c$ minimizes orbital contributions.
The main conclusions presented below apply to the entire sample series.

In general, microwave absorption can modify the longitudinal resistance through two conceptually distinct mechanisms.
The first is resonant heating of the electron system, for which the electrically detected ESR signal is expected to
scale as $dR/dT$. This mechanism is known to dominate ESR detection in the quantum Hall
regime~\cite{Shchepetilnikov-2021-PRB,Shchepetilnikov-2023-PRB}. Alternatively, microwave absorption can induce a
resonant change in the spin polarization, producing a resistance response governed by the spin sensitivity $dR/d\xi$.

\begin{figure}
\includegraphics[width=1\columnwidth,clip]{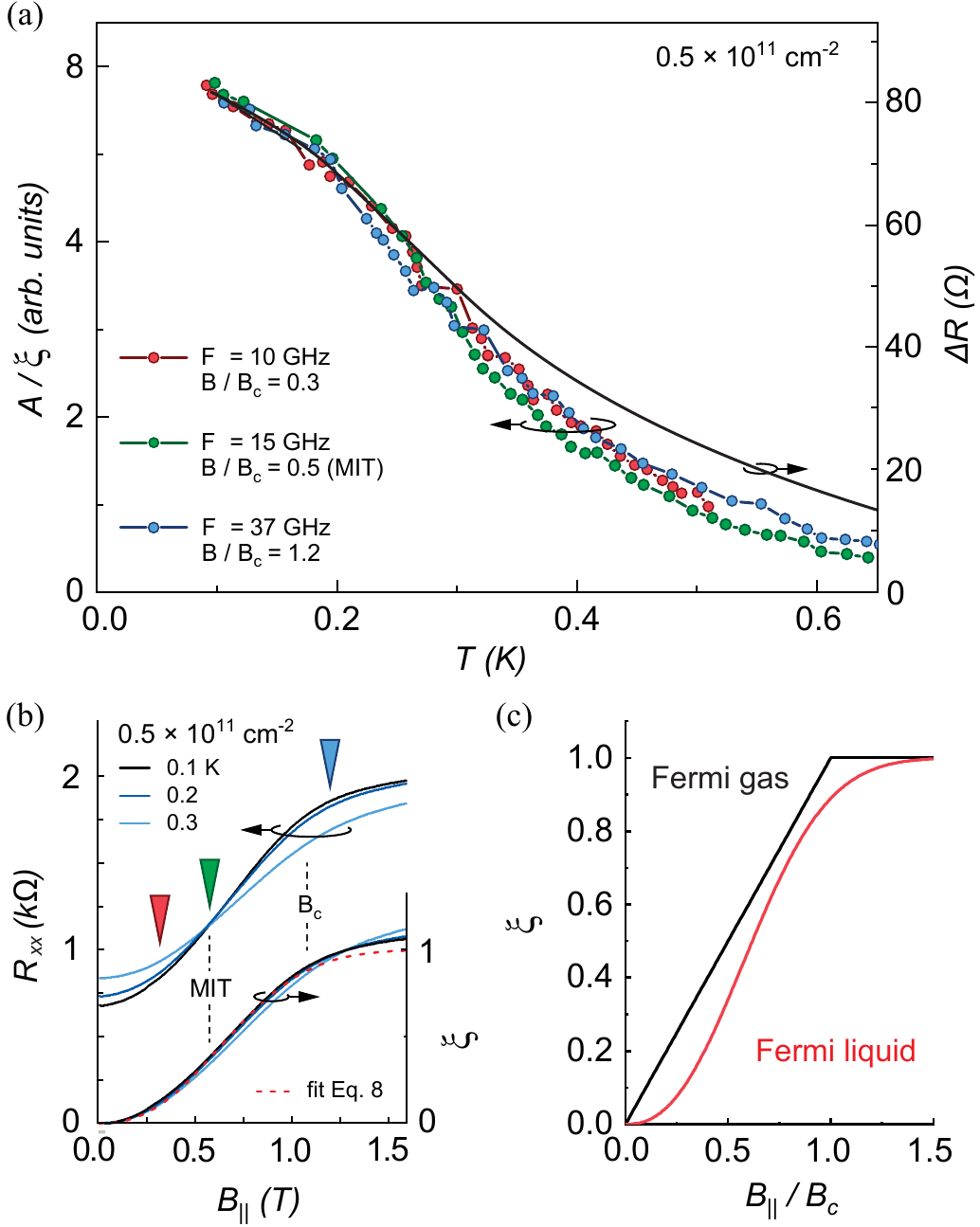}
\caption{
    (a) Temperature dependence of the ESR amplitude normalized by the spin polarization
    at $B_{\parallel}/B_c=0.3$, $0.5$, and $1.2$ (excitation frequencies of $10$, $15$, and $37$~GHz, respectively)
    for the sample with $n=0.5\times10^{11}$~cm$^{-2}$. The magnetic field $B_{\parallel}=0.5B_c$ coincides with the metal--insulator
    transition, where $dR_{xx}/dT=0$. The persistence of nearly identical temperature dependences at all three
    fields demonstrates that the ESR signal is governed by the spin sensitivity of the resistance rather than by resonant heating.
    The black curve shows $\Delta R \propto R_{xx}(0.2 B_c,T)-R_{xx}(0,T)$, in agreement
    with Eqs.~(\ref{eq:2}) and~(\ref{eq:3}).
    (b) Longitudinal resistance as a function of parallel magnetic
    field measured at three temperatures. The right axis shows the spin polarization reconstructed
    from Eq.~(\ref{eq:5}), demonstrating the collapse of data obtained at different temperatures.
    The dashed curve is the fit of Eq.~(\ref{eq:polarization_fit}).
    (c) Quasi-2D Fermi-liquid fit using Eq.~(\ref{eq:polarization_fit}) (red line), compared with the prediction for a non-interacting 2D Fermi gas (black line).
}\label{fig2}
\end{figure}

These two detection mechanisms can be distinguished by analyzing the temperature dependence of the ESR amplitude normalized by the spin polarization, $A/\xi(B)$.
This normalization removes the known contribution of spin polarization to the ESR intensity~\cite{Shchepetilnikov-2024-PRL,Shchepetilnikov-2024-JETPLetters}.
For the initial normalization the polarization was estimated using the non-interacting
relation $\xi=B_{\parallel}/B_c$ and was subsequently replaced by the experimentally
reconstructed $\xi(B)$ introduced later in the manuscript. This self-consistent correction did not alter the conclusions.
The comparison is performed at three representative fields, $B_{\parallel}/B_c=0.3$, $0.5$, and $1.2$ (triangles in Fig.~\ref{fig2}b),
corresponding to excitation frequencies of $10$, $15$, and $37$~GHz, respectively (Fig.~\ref{fig2}a).
Remarkably, the intermediate field coincides with
the metal--insulator transition, where $dR/dT=0$ and any bolometric contribution is expected to vanish~\cite{Shashkin-2019-MDPI}. Nevertheless,
the ESR amplitude exhibits nearly identical temperature dependences at all three fields. The persistence of a strong
resonance signal at the metal--insulator transition therefore excludes resonant heating as the dominant detection
mechanism and demonstrates that, in the absence of Landau quantization, the electrically detected ESR signal is
governed primarily by changes in spin polarization (for more details, see Supplemental Material~III).
Accordingly, the resonance amplitude is described by
\begin{equation}
\label{eq:1}
A(B,T) \sim  I\,\xi\,\frac{\partial R}{\partial\xi}(B,T),
\end{equation}
where $I$ is the incident radiation intensity.

A key observation is that the temperature evolution of $A/\xi \sim \partial R/\partial\xi$ is
independent of the resonance field within experimental uncertainty. Equation~(\ref{eq:1}) therefore implies
\begin{equation}
\label{eq:2}
\frac{\partial R}{\partial\xi}(B,T)
=
\frac{\partial R}{\partial\xi}(T),
\end{equation}

This allows integration between the unpolarized state and an arbitrary spin polarization, yielding
\begin{equation}
\label{eq:3}
R(B,T)-R_0(T)
=
\xi(B)\,
\frac{\partial R}{\partial\xi}(T),
\end{equation}

where $R_0(T)\equiv R(0,T)$.
Equation~(\ref{eq:3}) is valid at arbitrary magnetic fields, including in the fully polarized state, $\xi=1$, and hence

\begin{equation} \label{eq:4}
\frac{\partial R}{\partial\xi}(T)
=
R_1(T)-R_0(T),
\end{equation}

where $R_1(T)\equiv R(B_1,T)$ denotes the resistance in the fully spin-polarized state.
Because the crossover to complete spin polarization is continuous rather than abrupt, we take $B_1\simeq1.2B_c$,
immediately beyond the magnetoresistance kink. Within the experimental uncertainty in determining $B_c$,
this definition is consistent with the conventional criterion that associates complete spin polarization with $B_c$~\cite{Vitkalov-2000-PRL, Falson-2022-NatureMaterials}.

In our samples, orbital effects become increasingly pronounced at high
magnetic fields and can modify the longitudinal resistance independently of the spin polarization. To minimize these
contributions, it is therefore advantageous to evaluate Eq.~(\ref{eq:3}) in the low-field limit.
Accordingly, the black curve in Fig.~\ref{fig2}a shows the quantity
$\partial R/\partial\xi(T) \sim R(0.2 B_c,T)-R(0,T)$. The good agreement with the
measured temperature dependence of the ESR intensity demonstrates the validity of Eqs.~(\ref{eq:1}), (\ref{eq:2})  and~(\ref{eq:3}).

Combining Eqs.~\ref{eq:3} and~\ref{eq:4}, we obtain

\begin{equation}
\label{eq:5}
\xi(B)=
\frac{R(B,T)-R_0(T)}
     {R_1(T)-R_0(T)}.
\end{equation}

Equation~(\ref{eq:5}) is the central result of this work. It establishes a direct, parameter-free relation between
the longitudinal resistance and the spin polarization of a strongly interacting two-dimensional electron system.
ESR provides an independent validation of this correspondence. Once established, the relation allows the
spin polarization to be reconstructed from conventional magnetotransport alone. Thus, $R(B,T)$ provides an experimentally accessible
electrical measure of the magnetic state.

To validate Eq.~(\ref{eq:5}), we compare the measured ESR amplitude $A(B, T)$ with the spin polarization reconstructed independently from magnetotransport.
The resulting polarization, shown by the solid curves in Fig.~\ref{fig1}f, is obtained without adjustable parameters.
Over the range where orbital
effects remain weak, these curves closely reproduce the measured ESR intensity across two orders of magnitude.
Values formally exceeding $\xi=1$ do not imply overpolarization. Rather, they reflect additional orbital
magnetoresistance that lies outside the spin-only description~\cite{DasSarma-2000-PRL}.

Another independent test is presented in Fig.~\ref{fig2}b.
Although the measured magnetoresistance varies appreciably between $0.1$, $0.2$, and $0.3$~K, the polarization
extracted using Eq.~(\ref{eq:5}) collapses onto a single curve. This collapse demonstrates that the apparent temperature
dependence of the magnetoresistance is contained entirely in $R(T)$, $R_0(T)$ and $R_1(T)$, whereas $\xi(B)$ remains unchanged.
Deviations emerge only above $B_c$, where orbital contributions cause the resistance to continue increasing after
complete spin polarization has been reached.

\begin{figure}
\includegraphics[width=1\columnwidth,clip]{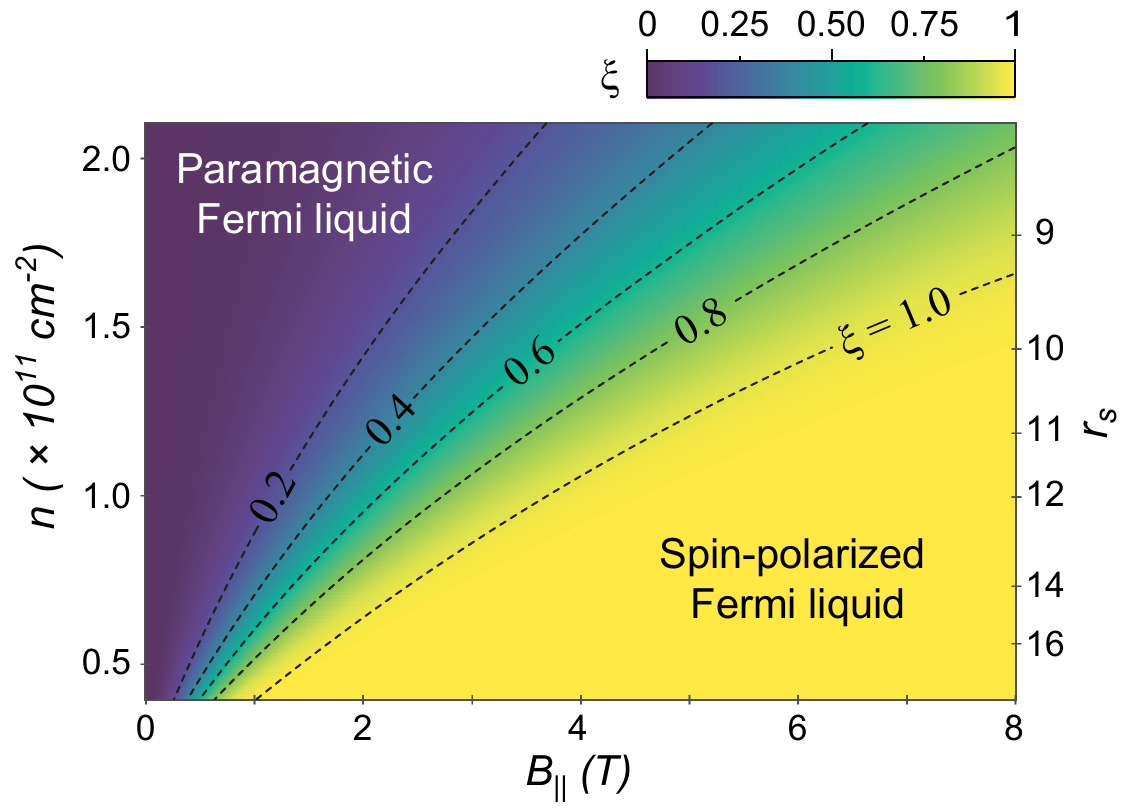}
\caption{
    Spin polarization reconstructed from magnetotransport using Eq.~(\ref{eq:polarization_fit}) is shown as a function of parallel magnetic field and electron density.
    The color scale represents the spin polarization $\xi(B, n)$, while dashed contours denote constant polarization.
    The diagram visualizes the continuous evolution of the electron liquid from a partially spin-polarized paramagnetic
    Fermi liquid to a fully spin-polarized Fermi liquid.
    The right axis shows the corresponding interaction parameter (dimensionless Wigner-Seitz radius)
    $r_s \equiv 1 / (\sqrt{\pi n} a_B)$, where effective Bohr radius is $a_B = 4 \pi \varepsilon_0 \varepsilon \hbar^2 / (e^2 m^*)$.
}\label{fig3}
\end{figure}

Finally, we find that the magnetic-field dependence of the spin polarization reconstructed from magnetotransport using Eq.~(\ref{eq:5})
is captured by the empirical relation

\begin{equation}
\label{eq:polarization_fit}
\xi(B,n)=
1-\exp\left[
-a\left(\frac{B}{B_c(n)}\right)^k
\right].
\end{equation}

This equation universally describes the spin polarization reconstructed using Eq.~\ref{eq:5} from published
magnetotransport data for Si MOSFETs~\cite{Vitkalov-2000-PRL}, SiGe/Si~\cite{Okamoto-2004-PRB}, AlAs/AlGaAs~\cite{Hossain-2020-PNAS},
ZnO/MgZnO~\cite{Falson-2022-NatureMaterials} heterostructures, and layered semiconductors such as MoTe$_2$ and MoSe$_2$~\cite{YangXu-2025-Arxiv,Tutuc-2018-PRB}.
For the ZnO/MgZnO samples investigated here,
the best-fit parameters are $a\simeq2.2$ and $k\simeq2.4$. Minor deviations appear only in the vicinity of $B_c$,
where orbital effects become appreciable, as illustrated by the red dashed curve in Fig.~\ref{fig2}b.

Figure~\ref{fig2}c compares the spin polarization reconstructed using Eq.~(\ref{eq:polarization_fit}) with the prediction
for a non-interacting two-dimensional Fermi gas. The pronounced nonlinear evolution
demonstrates that the magnetic response of the interacting electron liquid cannot be understood simply in terms of the
redistribution of electrons between two Zeeman-split spin parabolas.

Having established that the spin polarization can be determined using Eq.~(\ref{eq:polarization_fit}),
we reconstruct the magnetic state of the two-dimensional electron liquid throughout the experimentally
accessible parameter space. Figure~\ref{fig3} presents the resulting spin-polarization map as a function of electron density
and in-plane magnetic field. This experimentally reconstructed spin-polarization map directly visualizes the magnetic evolution of the Fermi-liquid state
over a broad density range, while remaining consistent with previous studies of ZnO/MgZnO heterostructures~\cite{Falson-2022-NatureMaterials},
which showed that at still lower electron densities the system evolves beyond the Fermi-liquid regime into more strongly correlated
electronic phases.

In summary, we have established a direct quantitative relation between the longitudinal resistance and the spin polarization
of a two-dimensional electron liquid. Using electrically detected electron spin resonance as an
independent calibration of the spin state, we demonstrate that the spin polarization can be reconstructed from conventional magnetotransport
measurements through a parameter-free relation (Eq.~\ref{eq:5}). The reconstructed polarization quantitatively reproduces the magnetic-field
and temperature dependence of the ESR response and reveals a strongly nonlinear evolution with magnetic field (Eq.~\ref{eq:polarization_fit}), reflecting
the effects of electron--electron interactions. This correspondence enables the magnetic state of the electron liquid to
be mapped over a broad range of carrier densities and magnetic fields using conventional transport measurements alone.

\textit{Acknowledgments}---The work has been supported by the Russian Federation Ministry of Science and
Higher Education as part of the state assignment of Osipyan Institute of Solid State Physics, Russian Academy of Sciences.
We thank J. Falson for the ZnO/MgZnO samples.
\nocite{*}
\putbib[references-main]
\end{bibunit}

\clearpage
\onecolumngrid
\begin{center}
\textbf{\large Supplemental Material for "Spin Polarization of a Two-Dimensional Electron Liquid"}
\end{center}

\vspace{0.5cm}

\noindent
G.~A.~Nikolaev,$^{1,2}$
M.~E.~Sergeev,$^{1,2}$
I.~V.~Kukushkin,$^{1}$
and A.~V.~Shchepetilnikov$^{1,3}$

\vspace{0.3cm}

\noindent
{\small
$^{1}$Osipyan Institute of Solid State Physics,
Russian Academy of Sciences,
Chernogolovka, Moscow Region 142432, Russia\\
$^{2}$Moscow Institute of Physics and Technology,
Dolgoprudny, Moscow Region 141701, Russia\\
$^{3}$National Research University Higher School of Economics,
101000 Moscow, Russia\\
}

\vspace{0.5cm}

\twocolumngrid
\begin{bibunit}[apsrev4-2]
\setcounter{equation}{0}
\setcounter{figure}{0}
\setcounter{table}{0}
\setcounter{page}{1}
\thispagestyle{empty}
\renewcommand{\thefigure}{S\arabic{figure}}

\begin{center}
\textbf{\large I. Method and samples}
\end{center}

The experiments were performed on four high-mobility ZnO/MgZnO two-dimensional electron systems with electron densities
of $n=0.5$, $1.2$, $1.4$, and $1.8\times10^{11}$~cm$^{-2}$ grown by molecular-beam epitaxy~\cite{Falson-2018-IOP}.
Three samples were patterned into van der Pauw geometries with soldered indium Ohmic contacts ($n=0.5$, $1.4$, and $1.8\times10^{11}$~cm$^{-2}$),
whereas the $n=1.2\times10^{11}$~cm$^{-2}$ sample was fabricated in a Corbino geometry.
The corresponding low-temperature electron mobilities ranged from $1\times10^{5}$ to $7\times10^{5}$~cm$^{2}$\,V$^{-1}$\,s$^{-1}$.
Measurements on the $n=1.2$, $1.4$, and $1.8\times10^{11}$~cm$^{-2}$ samples were carried out in a liquid-$^4$He cryostat
at temperatures down to $1.5$~K and magnetic fields up to $15$~T, whereas the lowest-density sample was studied
in a dilution refrigerator at temperatures down to $100$~mK and magnetic fields up to $14$~T.

We next describe the experimental scheme used to detect electron spin resonance.
The transport response was measured using a double lock-in technique to enhance the signal-to-noise ratio~\cite{Shchepetilnikov-2021-PRB,Shchepetilnikov-2024-PRL}.
The first lock-in amplifier measured the sample resistance $R_{xx}$ using an AC current at $1147$~Hz.
The incident electromagnetic radiation was amplitude-modulated at $23$~Hz,
and its absorption induced an oscillating resistance variation $\delta R_{xx}$,
detected by the second lock-in amplifier. Electron spin resonance appeared as a sharp peak in $\delta R_{xx}$
upon sweeping the magnetic field at fixed radiation frequency. The resonance shape and amplitude were independent
of the modulation frequencies and magnetic-field sweep rate.

\begin{center}
\textbf{\large II. Illumination intensity normalization}
\end{center}

\begin{figure*}
\includegraphics[width=2\columnwidth,clip]{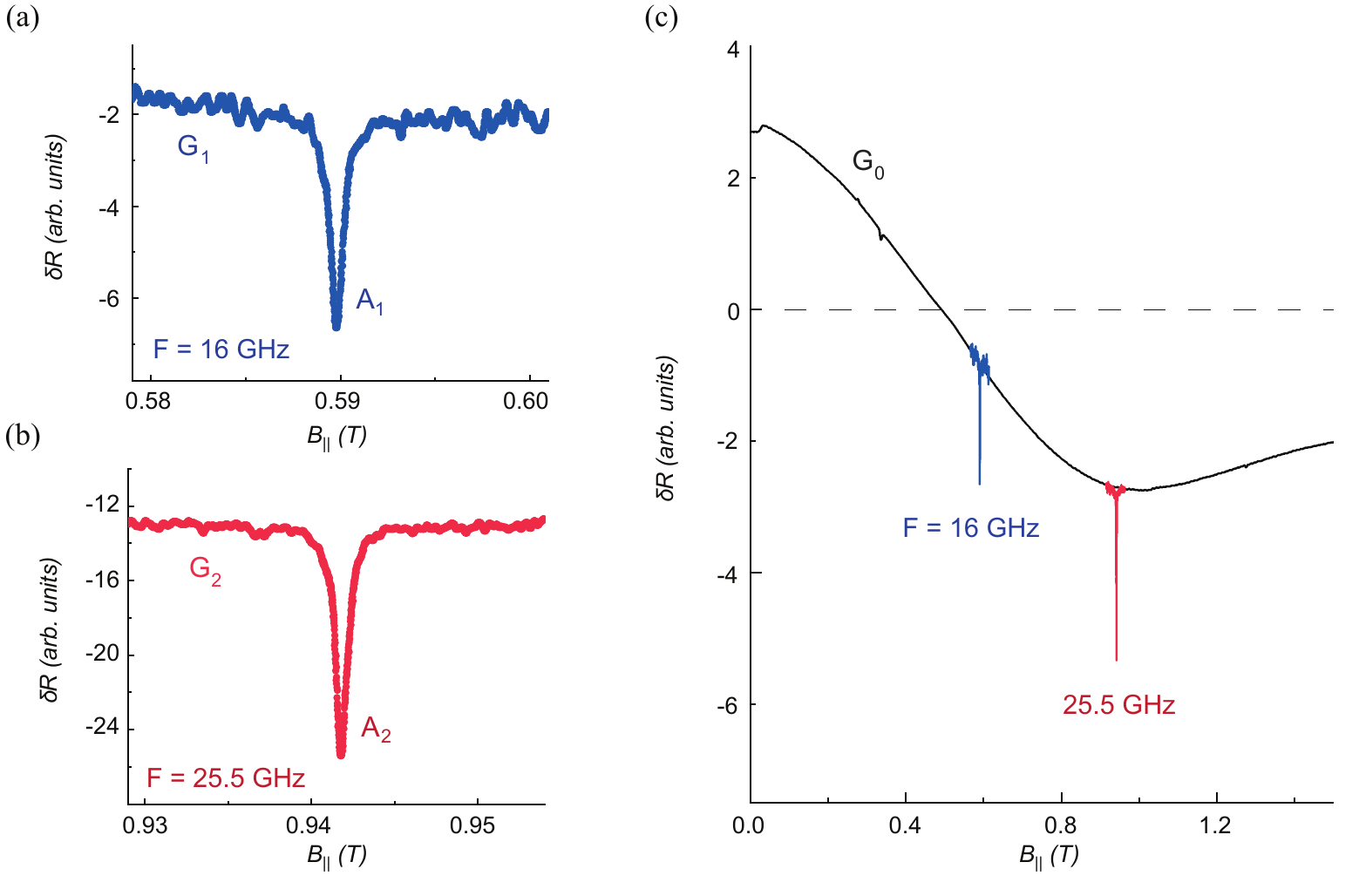}
\caption{
    (a),(b) Representative electrically detected ESR signals, $\delta R$, measured at two excitation frequencies on a ZnO/MgZnO heterojunction
    with electron density $n=0.5\times10^{11}$~cm$^{-2}$.
    The measured signal consists of a resonant ESR contribution ($A_i$) superimposed on a non-resonant microwave background ($G_i$), both proportional to the incident microwave intensity.
    (c) A reference background $G_0$ measured at $F = 9$~GHz over a wide magnetic-field range is used to normalize ESR signals $\delta R$
    recorded at different microwave intensities. After rescaling to the common background, $G_0$ is subtracted and the
    integrated ESR intensity is extracted. This procedure removes the trivial frequency dependence of the microwave
    intensity while preserving the intrinsic ESR response, allowing quantitative comparison of ESR signals $\delta R$
    measured under different experimental conditions.
}\label{fig:S1}
\end{figure*}

Broadband ESR measurements were performed over a wide range of excitation frequencies.
Because the microwave transmission of the experimental setup depends on frequency,
the microwave intensity reaching the sample varies from one frequency to another.
Consequently, the amplitudes of electrically detected ESR signals $\delta R$ measured at different frequencies cannot be compared directly.
Fig.~\ref{fig:S1} summarizes the normalization procedure used to place all signals $\delta R$ on a common microwave-intensity scale.

Representative ESR signals $\delta R$ measured on a ZnO/MgZnO heterojunction with electron density $n=0.5\times10^{11}$~cm$^{-2}$
at excitation frequencies of $F=16$ and $25.5$~GHz are shown in Figs.~\ref{fig:S1}a,b.
Each measured trace consists of a resonant ESR contribution superimposed on a smooth non-resonant microwave background,

\begin{equation}
\delta R(B,T)=A(B,T)+G(B,T),
\label{eq:A_B10}
\end{equation}

where $A(B,T)$ is the resonant ESR response and $G(B,T)$ is the non-resonant microwave background.

Throughout this work, all measurements were performed in the linear-response regime.
The amplitudes of both the resonant ESR signal and the non-resonant background were verified experimentally to be
proportional to the incident microwave intensity over the entire range of excitation powers employed in this work.

Accordingly,

\begin{equation}
\begin{aligned}
A(B,T)&=
I\,a(B,T),\\
G(B,T)&=
I\,g(B,T),
\end{aligned}
\label{eq:A_B11}
\end{equation}

where $I$ is the microwave intensity at the sample position. The function $g(B,T)$ describes the non-resonant
microwave background, whereas $a(B,T)$ characterizes the intrinsic electrically detected ESR response.
At this stage no assumption is made regarding the explicit form of $a(B,T)$ and $g(B,T)$.

To compare ESR signals $\delta R$ measured at different excitation frequencies, a reference background

\[
G_0(B,T)=I_0g(B,T)
\]

is first recorded over a wide magnetic-field range using a fixed microwave intensity $I_0$ (Fig.~\ref{fig:S1}c).
Since the background scales linearly with microwave intensity, $G_0(B,T)$ provides a common reference against
which all ESR signals $\delta R$ are normalized. Consider two measurements performed at microwave intensities $I_1$ and $I_2$,

\begin{equation}
\begin{cases}
\delta R_1(B_1,T)=A_1(B_1,T)+G_1(B_1,T),\\
\delta R_2(B_2,T)=A_2(B_2,T)+G_2(B_2,T).
\end{cases}
\label{eq:A_B12}
\end{equation}

Multiplying the signals $\delta R$ by the corresponding factor $I_0/I_i$ gives

\begin{equation}
\begin{cases}
\dfrac{I_0}{I_1}\delta R_1(B_1,T)=
\dfrac{I_0}{I_1}A_1(B_1,T)+G_0(B_1,T),\\[0.4cm]
\dfrac{I_0}{I_2}\delta R_2(B_2,T)=
\dfrac{I_0}{I_2}A_2(B_2,T)+G_0(B_2,T),
\end{cases}
\label{eq:A_B13}
\end{equation}

so that all normalized ESR signals share exactly the same non-resonant background.
After subtracting the common reference background $G_0(B_i,T)$, the remaining resonant ESR signals become

\begin{equation}
\begin{cases}
\dfrac{I_0}{I_1}A_1(B_1,T)=
I_0a(B_1,T),\\[0.3cm]
\dfrac{I_0}{I_2}A_2(B_2,T)=
I_0a(B_2,T).
\end{cases}
\label{eq:A_B14}
\end{equation}

The normalization therefore removes the trivial variation of microwave intensity between different excitation
frequencies while preserving the intrinsic ESR response of the two-dimensional electron system. Consequently,
all ESR amplitudes are placed on a common absolute scale and can be compared quantitatively.

Finally, the ESR response is characterized by its integrated intensity rather than by its peak amplitude.
As illustrated in Fig.~\ref{fig:S1}c, the integrated intensity is determined after subtraction of the common reference background.
Compared with the peak height, the integrated intensity is considerably less sensitive to frequency-dependent
variations of the resonance linewidth and lineshape, which may substantially modify the peak amplitude while
leaving the total absorbed microwave power nearly unchanged~\cite{Nikolaev-2026-APL}.
The typical measurement error of the integrated ESR amplitude is approximately $10\%$, arising primarily from the intensity normalization procedure.

The explicit form of the function $a(B,T)$ is established in the main text and Supplemental Material~III, where we show that

\begin{equation}
a(B,T)=
\xi(B)\,
\frac{\partial R}{\partial\xi}(T).
\end{equation}

The normalized ESR amplitude therefore factorizes into independent magnetic-field and temperature contributions.
At fixed magnetic field it is proportional to $\partial R/\partial\xi(T)$, whereas at fixed temperature it is proportional
to the equilibrium spin polarization $\xi(B)$. This separation forms the basis of the analysis presented in the main text.

\begin{center}
\textbf{\large III. Electrically detected spin resonance spectroscopy}
\end{center}

The normalization procedure described in Supplemental Material~II places all ESR signals $\delta R$ on a common microwave-intensity scale,
allowing the resonant ESR response to be written in the general form

\begin{equation}
A(B,T)=Ia(B,T),
\label{eq:A_B1}
\end{equation}

where $a(B,T)$ characterizes the intrinsic electrically detected ESR response.
In this section, we determine the explicit form of the function $a(B,T)$ for the case of a magnetic field applied parallel to the electron system.

For sufficiently weak microwave excitation, and in the absence of orbital effects, the microwave-induced change in the longitudinal resistance
can be expressed to first order in terms of variations in the electron temperature and spin polarization,

\begin{equation}
\delta R(B,T)
=
\Delta T
\frac{\partial R}{\partial T}(B,T)
+
\Delta\xi
\frac{\partial R}{\partial\xi}(B,T).
\label{eq:A_B2}
\end{equation}

Equation~(\ref{eq:A_B2}) describes the response of the electron system to an arbitrary perturbation.
Throughout this work, all measurements were performed in the linear-response regime, where the microwave-induced
resistance change is proportional to the incident microwave intensity. The linear dependence of both the resonant
ESR signal and the non-resonant microwave background on microwave power was verified experimentally over the full
range of excitation powers employed in this study. The measured response can therefore be expressed as

\begin{equation}
\delta R(B,T)
=
I
\left(
a(B,T)
+
g(B,T)
\right),
\label{eq:A_B3}
\end{equation}

where $I$ is the microwave intensity at the sample position, $a(B,T)$ describes the resonant ESR response,
and $g(B,T)$ denotes the non-resonant microwave background (for details, see Supplemental Material~II).

Previous studies established experimentally that the resonant electrically detected ESR amplitude is proportional to
the equilibrium spin polarization in strongly correlated two-dimensional electron systems~\cite{Shchepetilnikov-2024-PRL,Shchepetilnikov-2024-JETPLetters},

\begin{equation}
A(B)\propto\xi(B).
\label{eq:A_xi}
\end{equation}

This proportionality can be understood from the limiting case of a non-interacting two-level system,
where the net ESR absorption is determined by the difference between the probabilities of photon absorption,
$\omega_{\rm abs}\propto wn_\uparrow(n-n_\downarrow)$,
and stimulated emission,
$\omega_{\rm em}\propto wn_\downarrow(n-n_\uparrow)$.
Here $w$ is the transition probability and $n$ is the degeneracy of each spin subband. Their difference satisfies

\[
A
\propto
\omega_{\rm abs}-\omega_{\rm em}
\propto
n_\uparrow-n_\downarrow
\propto
\xi,
\]

This simple two-level picture illustrates why the ESR amplitude is expected to scale with the equilibrium spin polarization.

Combining Eqs.~(\ref{eq:A_B2}), (\ref{eq:A_B3}) and (\ref{eq:A_xi}), the linear-response expression for the resonant ESR signal becomes

\begin{equation}
A(B,T)
=
I\xi(B)
\left(
\alpha
\frac{\partial R}{\partial T}(B,T)
+
\beta
\frac{\partial R}{\partial\xi}(B,T)
\right),
\label{eq:A_B4}
\end{equation}

where $\alpha$ and $\beta$ are phenomenological coefficients describing the thermal and spin-polarization contributions
to the electrically detected ESR signal, respectively.
In the quantum Hall regime, the electrically detected ESR response is frequently dominated by the thermal contribution,
as observed in ZnO/MgZnO~\cite{Shchepetilnikov-2023-PRB,Shchepetilnikov-2024-JETPLetters}
and GaAs/AlGaAs~\cite{Olshanetsky-2003-PRB}. Nevertheless, spin-dependent detection ($\beta\neq0$) has also been demonstrated under
quantum Hall conditions, for example in Si/SiGe quantum wells~\cite{Matsunami-2006-PRL}. Understanding the factors governing the
relative importance of these detection mechanisms remains an important challenge.

The dominant detection mechanism in a parallel magnetic field can be identified experimentally from the temperature dependence of the normalized ESR amplitude.
After normalization by the equilibrium spin polarization, the ESR amplitudes $A/\xi$ measured at $B_{\parallel}/B_c=0.3$, $0.5$, and $1.2$ exhibit
nearly identical temperature dependences (Fig.~2a of the main text), including at the metal--insulator transition, $B_{\parallel}=0.5B_c$, where

\[
\frac{\partial R}{\partial T}(0.5B_c,T)=0.
\]

The persistence of a pronounced ESR signal under these conditions excludes resonant heating as the dominant detection mechanism.
Consequently, $\alpha\simeq0$, and Eq.~(\ref{eq:A_B4}) reduces to

\begin{equation}
A(B,T)
\propto
I\xi(B)
\frac{\partial R}{\partial\xi}(B,T).
\label{eq:A_B5}
\end{equation}

Experimentally, the normalized ESR amplitude,

\[
\frac{A(B,T)}{\xi(B)},
\]

exhibits nearly identical temperature dependences at all investigated magnetic fields (Fig.~2a of the main text).
More generally, the observed collapse of the normalized ESR amplitudes $A/\xi$ implies the factorization

\begin{equation}
\frac{A(B,T)}{\xi(B)}
=
f_1(B)f_2(T),
\end{equation}

where $f_1(B)$ and $f_2(T)$ are independent functions of magnetic field and temperature.
Experimentally, the normalized ESR amplitude, $A/\xi$, exhibits nearly identical temperature dependences at all investigated
magnetic fields (Fig.~2a of the main text). This observation implies that the factor $f_1(B)$ is approximately unity
within the experimental uncertainty, indicating that the magnetic-field dependence of the ESR amplitude is entirely
contained in the equilibrium spin polarization.
Comparison with Eq.~(\ref{eq:A_B5}) therefore shows that, to an excellent approximation,

\begin{equation}
\frac{\partial R}{\partial\xi}(B,T)
=
\frac{\partial R}{\partial\xi}(T).
\label{eq:A_B6}
\end{equation}

This is the central statement used in deriving Eq.~(7) of the main text.
\putbib[references-supplementary]
\end{bibunit}

\end{document}